\documentclass[9pt,twoside]{extarticle}

\usepackage[margin=0.75in, top=0.75in, bottom=0.75in,
            columnsep=0.25in]{geometry}
\usepackage{multicol}

\usepackage[T1]{fontenc}
\usepackage{mathptmx}
\usepackage[scaled=0.86]{helvet}
\usepackage{microtype}

\usepackage{amsmath,amssymb,bm}
\usepackage{mathtools}

\usepackage{float}

\usepackage{graphicx}
\usepackage{booktabs}
\usepackage{dcolumn}
\usepackage[rightcaption]{sidecap}
\usepackage[labelfont=bf, 
            font=small,
            labelsep=period]{caption}
\usepackage{enumitem}

\usepackage{engord}

\usepackage{xcolor}
\usepackage{url}
\newcommand{\doi}[1]{\href{https://doi.org/#1}{\detokenize{#1}}}
\usepackage{hyperref}
\usepackage{hycolor}
\hypersetup{colorlinks=true,citecolor=[rgb]{0,0.18,0.68}, 
urlcolor=[rgb]{0,0.18,0.68}, linkcolor=[rgb]{0,0.18,0.68}, final = true}
\usepackage{relsize}

\usepackage[super,compress,sort]{natbib}
\setcitestyle{numbers,super,open={},close={}}

\usepackage{titlesec}
\titleformat{\section}{\bfseries\sffamily\uppercase}{\thesection.}{0.5em}{}
\titleformat{\subsection}{\bfseries\sffamily}{\thesubsection.}{0.5em}{}
\titleformat{\subsubsection}{\itshape}{\thesubsubsection.}{0.5em}{}
\titleformat{\paragraph}[runin]{\bfseries\small}{\theparagraph}{0.5em}{}

\titlespacing\subsection{0pt}{8pt plus 4pt minus 2pt}{2pt plus 2pt minus 2pt}
\titlespacing\subsubsection{0pt}{8pt plus 4pt minus 2pt}{0pt plus 2pt minus 2pt}
\titlespacing\paragraph{12pt}{4pt plus 4pt minus 2pt}{4pt plus 2pt minus 2pt}

\usepackage{fancyhdr}
\usepackage{abstract}

\usepackage{siunitx}
\titleformat{\section}
    {\fontsize{12}{10}\bfseries\sffamily\uppercase}
    {\thesection.}{0.5em}{}

  \def\be{\begin{equation*}}
  \def\ee{\end{equation*}}
  \def\ba{\begin{eqnarray}}
  \def\ea{\end{eqnarray}}
  
  \def\fref#1{Fig.~\ref{#1}}

  \def\bt{\textrm} 
  \def\nsb#1{\noindent\textbf{\bt{#1~}}}
  
  \definecolor{or}{RGB}{234,142,53}
  \definecolor{gr}{RGB}{150,150,150}
  \definecolor{bl}{RGB}{54,152,187}

  \newcommand{\ie}{\textit{i.e.}}
  \newcommand{\eg}{\textit{e.g.}}

  \definecolor{YKB}{rgb}{0.00,0.18,0.65}

  \def\Fti{F_{t,i}}                      
  \def\fti{f_{t,i}}                      
  \def\atti{a_{t+1,i}}                      
  \def\pti{p_{t,i}}                      
  \def\wti{w_{t,i}}                      
  \def\smax{s_{\mathrm{max}}}            
  \def\sval{s}                           
  \def\sgamma{\gamma}                    
  \def\sbeta{\beta}                      
  \def\sridge{\sigma}                    
  \def\daic{\Delta\mathrm{AIC}}

\begin{document}

\title{\LARGE\bfseries\sffamily Population Ecology of Tunes}

\author{
    John M. McBride\textsuperscript{1,2,*} 
    \and 
    Armand M. Leroi\textsuperscript{3,4}
}

\date{
    \small
    \textsuperscript{1}Department of Behavioral and Cognitive Biology, 
    University of Vienna, Vienna, Austria\\[0.5em]
    \textsuperscript{2}Acoustic Research Institute, Austrian Academy of Science,
    Vienna, Austria\\[0.5em]
    \textsuperscript{3}Department of Life Sciences, Imperial College London,
    London, UK\\[0.5em]
    \textsuperscript{4}Data Science Institute, Imperial College London,
    London, UK\\[0.5em]
    \textsuperscript{*}Correspondence: 
    \href{mailto:jmmcbride@protonmail.com}{jmmcbride@protonmail.com}\\
}

\twocolumn[
  \begin{@twocolumnfalse}
    \maketitle
  \end{@twocolumnfalse}
]

  \begin{abstract}
  How cultural repertoires maintain diversity under selection is a
  fundamental question in cultural evolution. We address this using
  thirteen years of weekly popularity data for approximately \num{20000} Irish traditional tunes, fitting ecological birth-process models under neutral, frequency-dependent, and per-tune selection hypotheses. We find strong evidence that tunes differ in intrinsic fitness -- some are systematically more likely to be learned than others.
  We find that \SI{29}{\%} of the variance in fitness can be explained
  by a mixture of social and melodic features. Some tunes appear to
  be carried along via linkage due to the tradition of playing tunes
  in sets, analogous to selective sweeps in genetics.
  By measuring changes in fitness over time and comparing this with
  recordings we precisely identify the mechanism by which a long-dormant tune
  can become fit through a popular recording.
  Despite the directional selection, repertoire diversity increases,
  driven by the continual arrival of new compositions. These results
  demonstrate that selection and diversity can coexist in a cultural
  ecosystem, and establish Irish traditional music as a quantitatively
  tractable system for studying the evolution of cultural variants
  and understanding what makes a tune stand out.
  \end{abstract}
  \vspace{0.5cm}
  \noindent\small\textbf{Keywords:} cultural evolution | cultural transmission | cultural diversity | 
  population ecology | folk music 
  \vspace{0.5cm}

\section*{Introduction} 

Living traditions maintain repertoires of cultural variants --- tunes, stories, dances, techniques --- that change over time. Some variants thrive and spread; others decline and are forgotten; new ones are continually created. This is fundamentally an ecological and evolutionary process, yet we rarely have the data to study it quantitatively over meaningful timescales. The central questions are those of any evolving system: is change neutral or selective? If selective, what determines relative fitness? And what maintains diversity in the face of selection? These questions have been explored theoretically in cultural evolution, where models of biased transmission, drift, and selection have been developed by analogy with population genetics\cite{cavallisforzaCultural1981,boydCulture1988,henrichEvolution1998}. Empirical tests have used frequency time series of cultural variants --- baby names, pottery motifs, dog breeds, pop charts, words, music samples --- to ask whether change is neutral or due to selection\cite{hahnDrift2003,bentleyRandom2004,bentleyRegular2007,shennanCeramic2001,herzogRandom2004,newberryDetecting2017,youngbloodConformity2019a}. Such data, however, are typically truncated to the most popular variants, aggregated over long intervals, or drawn from a population that cannot be delimited, which makes selection hard to distinguish from drift\cite{kandlerAnalysing2019,leroiNeutral2020}. What is needed is a system where the population is observable, the unit of transmission is well-defined, and the dynamics play out over a long enough period to distinguish selection from noise.

\begin{figure*}[ht!]
  \centering
  \includegraphics[width=\textwidth]{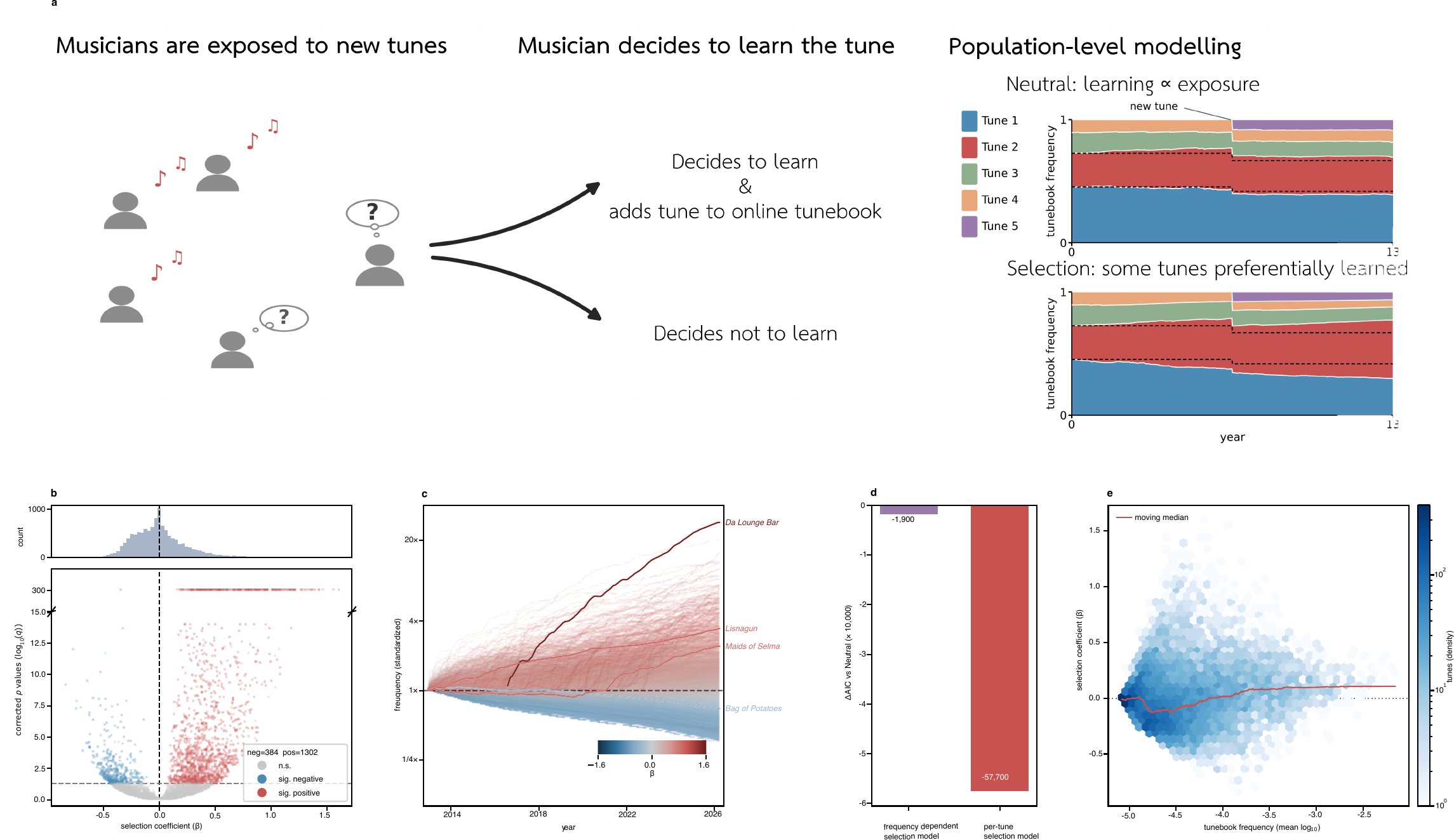}
  \caption{%
  \textbf{Inferring selection on tunes.} 
  \textbf{a.}~Schematic of tune transmission and how we model it. Musicians are exposed to tunes played by their peers, and decide whether or not to learn them; those who learn a tune may add it to
  their online tunebook. At the population level, we model each week's tunebook adds as draws from the previous week's distribution of tunebook entries across tunes (colours denote tunes; bar widths denote  proportions). Under neutrality, the probability that a tune is added  is proportional to its current popularity; under selection, some tunes are learned more (or less) often than their popularity predicts, and the distribution of adds deviates from the neutral expectation (dashed lines). 
  \textbf{b.}~Volcano plot of per-tune selection coefficients $\beta_i$. The upper histogram shows the marginal distribution of $\beta_i$; the main panel plots $-\log_{10}(q)$ (Benjamini-Hochberg corrected $p$-value) against $\beta_i$.
  Tunes with $q < 0.05$ are coloured red (positive) or blue (negative); grey points are non-significant.
  The dashed horizontal line marks the FDR threshold; the broken $y$-axis separates the dense main distribution from the small number of extremely significant tunes --- these share the same $q$ value due to limits on floating-point precision.
  \textbf{c.}~Evolution of relative tune frequencies in \texttt{tunebook}, 2012--2026, coloured by estimated per-tune selection coefficients, $\beta_i$. The grey line indicates neutrality; the trajectories of a few tunes are picked out in bold lines (see text).
  \textbf{d.}~$\daic$ relative to the neutral model for two selection models: frequency-dependent, and per-tune. 
  \textbf{e.}~Density of tunes by mean frequency (tunes whose \texttt{tunebook} count never reached 10 excluded). The red line shows a moving median of the per-tune selection coefficients, $\beta_i$, and weak selection against rare tunes indicating conformist bias.
  }
  \label{fig:fig1}
\end{figure*}

Irish traditional music offers an unusually good system for this purpose. The tradition is predominantly oral: tunes are learned by
ear, passed between players in social settings, and carried primarily in memory rather than on paper\cite{breathnachUse1986,cawleyBecoming2020}. Crucially, the music is played in \textit{sessions} --- informal gatherings where musicians play tunes together in sets of two or more\cite{osheaGetting2007}. A player starts a tune and others join in if they know it; those who do not know it listen, and may later learn it. This social structure constitutes a clear mechanism of cultural transmission: a musician encounters a tune, decides whether to learn it, and either adds it to their active repertoire or does not (\fref{fig:fig1}a). 

The tradition has also embraced modern technology. A vast catalogue of professional and amateur recordings exists, supplemented by online courses and thousands of instructional and performance videos on platforms like YouTube\cite{wardTechnology2019}. Most significantly, the development of ABC notation on Irish music discussion forums in 1993\cite{walshawAbcnotationcom} catalysed the creation of online tune repositories -- websites where enthusiasts collaboratively transcribed and archived the tradition. The largest of these platforms, \texttt{The Session}\cite{keithThesessionorg}, provides weekly data on which tunes users add to personal \texttt{tunebooks}, yielding a longitudinal record of tune popularity spanning over thirteen years and approximately 20,000 tunes. Data from this platform have previously been used to track tune popularity and to relate melodic complexity to
popularity\cite{streetRole2022}. This is a remarkably rich dataset for studying cultural dynamics, comparable in resolution to long-term ecological census data\cite{conditThirty2012}, but for a cultural ecosystem.

Cultural evolutionists distinguish several forces that can influence the dynamics, and hence diversity, of copied artefacts.  Most simply, they may be copied strictly in proportion to their presence in the population. This model assumes no selection and is therefore neutral. Alternatively, the probability of copying may increase faster than frequency (positive frequency dependence or conformity bias) or less (negative frequency dependence, anti-conformity bias or novelty bias). Prestige bias supposes that tunes are chosen on the basis of who plays them. These are all forms of social selection, however, artefacts may be also chosen for intrinsic properties such as beauty. In the absence of new variants, neutrality and most forms of selection will reduce artefact diversity over time in finite populations; only negative frequency dependence will maintain it rather as niches do in ecological communities.
  
To disentangle these forces, we model the evolution of tune popularity using methods from population ecology, fitting birth processes under three evolutionary models: a neutral one, where tune popularity grows in proportion to current frequency; a frequency-dependent selection one; and a model in which each tune has its own fitness. We reject neutral evolution and find only weak evidence of frequency dependence. Instead, the dynamics of tune popularity are best explained by a model in which each tune has a fitness of its own independent of its frequency. Thus the evolution of this musical ecosystem is dominated by directional selection.  Examining the correlates of relative fitness, we find that we can predict the fitness of a tune from its prominence on social media, as well as its intrinsic musical properties, though the variance in fitness explained by these features is modest (29\%). Finally, we show that tune diversity has increased monotonically over the last thirteen years in the face of selection's homogenizing force, and that this increase is due to the constant arrival of new compositions. In sum, we show how the forces that shape a musical tradition can be quantified much as those that shape the evolution of organic populations and communities have been.

\section*{Results} 

\subsection*{Selection on tunes}

The neutral model predicts that the number of new \texttt{tunebook} adds in a period of time is proportional to a tune's popularity in the previous time period. We began by comparing this model to one in which a selection coefficient, $\beta_i$, was estimated for each tune (\fref{fig:fig1}b). \num{10425} (\SI{47}{\%}) of our tunes never reached a \texttt{tunebook} count of 10 members and, for these, the standard errors of the $\beta_i$ are very large, so we excluded them here and in all subsequent analyses leaving us with \num{11791} tunes (see SI for analyses including all tunes). A model based on this retained set showed overwhelming support for directional selection over neutrality ($\daic = -57700$). Selection coefficients ranged from $-0.84$ to $1.60$ (\fref{fig:fig1}b) which, over the thirteen years of \texttt{tunebook}'s evolution, had large effects on frequencies (\fref{fig:fig1}c). For example, \emph{Maids of Selma} (tune \texttt{2766}) began with 37 \texttt{tunebook} entries and ended with 260, while \emph{The Bag of Potatoes} (tune \texttt{391}) fell from \engordnumber{321} to \engordnumber{501} in the popularity ranking.

Given the prominence of conformist models in cultural evolution\cite{boydCulture1988,henrichEvolution1998}, we next built a frequency dependent selection model in which we allowed selection to vary as a global parameter, $\beta_f$, for all tunes, and found that it performs better than a neutral model ($\daic = -1000$) implying weak conformist selection ($\beta_f = 0.024$). Consistent with this, a plot of $\beta_i$ against mean frequency (\fref{fig:fig1}e) shows that tunes with a mean \texttt{tunebook} count below about 60 have a small selective disadvantage, while more popular tunes have a small advantage. After modeling an additional saturation effect --- adds may be limited by the availability of members --- we see a modest increase to $\beta_f=0.048$ ($\daic = -1900$). Even so, the $\daic$ of the frequency-dependent model is only around \SI{3}{\%} of the directional selection model's. We conclude that the most important selective forces shaping \texttt{tunebook} dynamics are directional, and that the causes of a tune's success or failure are unique to it. We next ask what they are.

\begin{figure*}[ht!]
  \centering
  \includegraphics[width=10 cm]{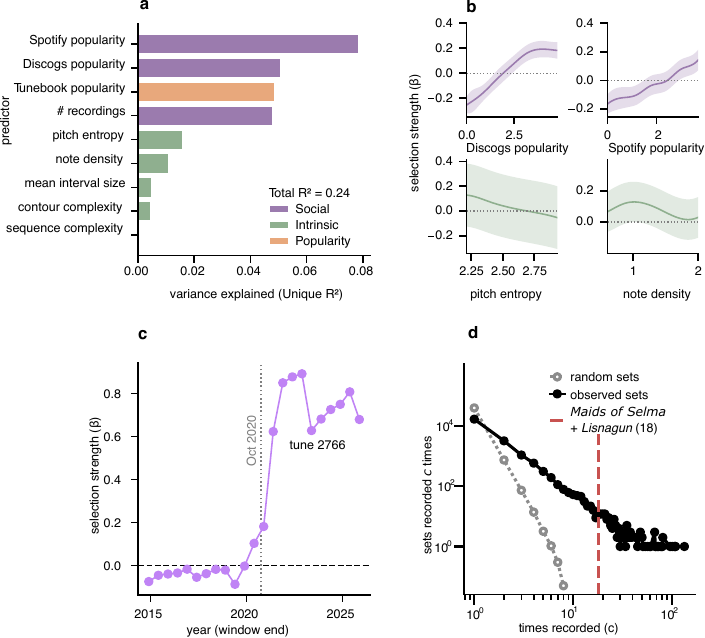}
  \caption{%
  \textbf{Predicting tune success.}
\textbf{a.}~Unique variance in $\beta_i$ explained by each predictor in a XGBoost model. Bars are coloured by predictor class: social popularity, \texttt{tunebook} popularity, intrinsic melodic. The legend reports the total cross-validated $R^2$.
\textbf{b.}~Partial effects of four predictors on $\beta_i$ (line: mean; band: \SI{95}{\%} CI), from a single multivariate GAM fitted on all nine predictors.
\textbf{c.}~Sliding-window estimate of $\beta$ for \emph{Maids of Selma} (\texttt{2766}), computed using two-year windows advanced in half-year steps; each estimate is plotted at the end of its window. The vertical dotted line marks October 2020, when a YouTube video featuring the tune was released.
\textbf{d.}~Sets of two tunes ranked by how often they are recorded in user-contributed sets (black points, solid line) relative to a null distribution in which sets are assembled at random according to tune popularity (grey points, dashed line). The red dashed line marks the set (\emph{Maids of Selma $+$ Lisnagun}) discussed in the text, recorded 18 times (rank 134 of \num{22982} distinct two-tune sets).}
  \label{fig:fig2}
\end{figure*}

\subsection*{Predicting tune success}

To understand the origin of differential fitness, we fitted an XGBoost regressor using two classes of features. Social variables capture how a tune is promoted through recordings: the number of recordings, representation on Discogs, and popularity on Spotify. To these we added an internal metric of popularity already discussed, the mean frequency of each tune. Melodic variables capture intrinsic properties of the tune itself: melodic pitch-class entropy, note density, contour complexity, and others (see Methods for the full list).

The XGBoost regressor achieved a Pearson correlation of $r = 0.56$ between out-of-fold predicted and observed $\beta_i$ (Supplementary Figure), corresponding to a held-out $R^2 = 0.29$ (5-fold cross-validation). External social popularity variables uniquely accounted for \SI{16}{\%} of the variance in $\beta_i$. Tune mean frequency -- reflecting the weak conformity bias identified in \fref{fig:fig1}d -- adds \SI{1}{\%} while melodic variables add another \SI{3}{\%}. These unique contributions account for about two-thirds of the total, so the three groups of variables carry largely non-overlapping information (\fref{fig:fig2}a).
 
Figure \ref{fig:fig2}b shows the effects of the most influential predictors as estimated by General Linear Models (GAMS). The strength of positive selection clearly increases with the two main prestige features: the number of recordings and maximum Spotify popularity across matched tracks. But the salutary effect of social media exposure on fitness can be illustrated most vividly by the fate of a particular tune.

\emph{Maids of Selma} (tune \texttt{2766}) is more than a hundred years old. Despite its age, it lay dormant in the \texttt{tunebook} for a long time: for six years it received only an average of 3.2 \texttt{tunebook} adds. We estimated the strength of selection, $\beta_i$, on this tune in moving windows of two years. Selection was slightly negative until around 2020, but then jumped to $\beta_i \geq 0.7$ (\fref{fig:fig2}c). This sudden rise in popularity appears to have been driven by a YouTube video released in October of that year and that has since accumulated more than 80,000 views. (We could not identify any contemporary recording that might have had a comparable effect.) Further analysis shows that, for every 500 views, there was approximately one new \texttt{tunebook} add (Supplementary Fig.~S14).

Such social effects outweigh the influence of our intrinsic musical features. Nevertheless, we found that selection becomes more negative as tunes become more complex as measured by chroma entropy and melodic note density (see Supplementary Fig.~S8 for related features). Simple tunes are, on average, more successful than complex ones. This result contrasts with the inverted-U relationship between melodic complexity and popularity reported previously for this dataset\cite{streetRole2022}.

All of our covariates are probably proxies for the true sources and targets of selection. Spotify popularity reflects the preferences of general listeners rather than the musicians who drive the oral transmission of our tunes; summary statistics like entropy and note density are likely only partially correlated with what matters most: the memorability and learnability of a tune. More fundamentally, our analysis assumed that a tune's fitness is determined only by its own properties, however, this need not be so since tunes are played in stereotyped \emph{sets}. It is to their effects that we now turn.

\subsection*{Musical linkage}

\texttt{TheSession.org} contains not only data on how often individual tunes are played but also how often they are played together. Some sets are so famous that they have been given names, e.g., The Coleman Set. We have \num{112642} such set records, \num{90634} of which contain two or more tunes (\num{56478} of them unique). We began by asking whether sets are assembled at random from tunes given their individual frequencies. Among the \num{41743} recorded sets of exactly two tunes, \num{22982} are distinct and the most common is recorded 134 times; when the same number of sets is assembled at random from tunes in proportion to their popularity, almost every set is unique (\num{40788} distinct on average) and no combination is recorded more than 8 times (\fref{fig:fig2}d). Thus sets are the result of deliberate choice.

Given this, the success of some tunes may depend on the success of others often played at the same time rather as the fitness of an allele may depend on others at linked loci. Several tightly linked genes may have a synergistic effect on each others' fitness (positive fitness epistasis within supergenes); but even unconditionally neutral or deleterious alleles may hitch-hike to high frequencies simply because they are linked to beneficial alleles \cite{bartonGenetic2000}. An analogous effect may, indeed, explain the \emph{Maids of Selma}'s sudden success. In the YouTube video that first brought this tune fame, it was paired --- to our knowledge, for the first time --- with another tune, \textit{Lisnagun} (tune \texttt{3842}). The two tunes now co-occur in 27 user-contributed sets (18 of them as an exact pair, ranked 134 of \num{22982} distinct pairs; \fref{fig:fig2}d), none of which predates the video. Since \emph{Lisnagun} had long been increasing in frequency ($\beta_i = 0.89$), it may be that \emph{Maids} is now just hitch-hiking on its partner's popularity.

\subsection*{Innovation dominates diversity}

\begin{figure}[htbp]
  \centering
  \includegraphics[width=\columnwidth]{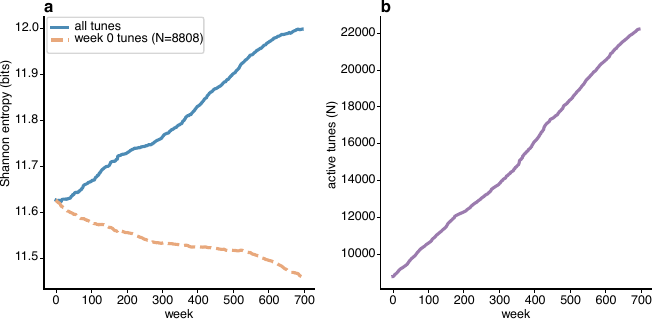}
  \caption{%
  \textbf{Ecosystem dynamics.} \textbf{a.}~Entropy of the popularity (cumulative \texttt{tunebook}-adds) distribution across all weeks for all tunes in the dataset (blue solid line), and for only the tunes that were already active in the first recorded week (orange dashed line).
  \textbf{b.}~Total number of active tunes --- tunes with a non-zero \texttt{tunebook} count --- by week.
  }
  \label{fig:fig3}
\end{figure}

Directional selection, if unopposed, should erode tune diversity over time as positively selected tunes increasingly dominate the repertoire. To test whether this is so, we tracked the entropy of the
popularity distribution across weeks, where higher entropy indicates a more even spread of tunebook adds across tunes (\fref{fig:fig3}a). We found that entropy in fact increases steadily, rising by \SI{0.37}{bits} over the observation period. This increase was entirely driven by the continual arrival of new tunes: the number of active tunes more than doubling, from roughly \num{9000} to over \num{22000} (\fref{fig:fig3}b). When we restricted the calculation to the \num{8808} tunes present at Week 0, we found that entropy gradually declines (\fref{fig:fig3}a), consistent with the concentration of popularity into positively selected tunes. Thus, while directional selection does erode the diversity of the repertoire, constant compositional innovation more than compensates, leading to a net increase in diversity. 
  
\section*{Discussion}

\nsb{Selection, diversity, and the role of technology.}
  The central finding of this study is that tune popularity in Irish
  traditional music is shaped by directional selection -- some tunes are
  systematically fitter than others -- yet the diversity of the
  repertoire is not eroding. This combination of selection and stable (or
  increasing) diversity is a familiar puzzle in ecology and evolutionary
  biology, and the mechanisms that resolve it there offer useful
  analogues here.

  The most direct explanation is the one supported by our data:
  innovation. New tunes enter the repertoire at a sufficient rate to
  offset the concentration of popularity into fitter existing tunes. This
  is analogous to the role of immigration and speciation in maintaining
  biodiversity under selection\cite{conditThirty2012}, and to the constant introduction of
  novelty that sustains diversity in cultural systems from baby names to
  pop charts\cite{bentleyRegular2007}. In our data, the number of active tunes more than doubles
  over the observation period, and this influx is a clear driver of the
  observed increase in repertoire entropy.

  Other mechanisms may also contribute, though we cannot test them
  directly. Niche structure is one: Irish traditional music encompasses
  distinct regional styles and tune types (reels, jigs, hornpipes,
  polkas, slides), and musicians may specialise, creating partially
  separated sub-communities within which different tunes are favoured.
  If these niches are stable, they can maintain diversity in the same way
  that spatial heterogeneity maintains species diversity in ecological
  communities\cite{chessonMechanisms2000}. Network structure is another candidate: the tradition is
  organised around local sessions, and tunes may spread primarily within
  clusters of musicians who play together regularly. If diffusion between
  clusters is slow, locally popular tunes can coexist even if one would
  dominate in a well-mixed population -- a dynamic analogous to
  population structure slowing the fixation of alleles in genetics\cite{whitlockFixation2003}.
  Fashion and temporal turnover may also play a role: the fitness of a
  tune need not be constant, and shifting tastes could prevent any single
  tune from dominating indefinitely. Our observation window is too short
  to detect long-period fluctuations in fitness, but the case studies
  hint at the possibility -- a tune can lie almost dormant 
  before experiencing a sudden resurgence.

  It is also worth noting that the distribution of fitness itself
  contributes to the maintenance of diversity. Most tunes have fitness
  values near zero: they are close to neutral, growing or declining only
  slightly faster than expected from their current popularity. Strong
  selection acts on the tails of the distribution, but for the bulk of
  the repertoire, stochastic dynamics -- who happens to play what at
  which session -- likely matter more than fitness differences. This is
  reminiscent of the nearly neutral theory of molecular evolution, in
  which most substitutions are too weakly selected to be amplified
  by selection, and drift dominates\cite{ohtaNearly1992}. In such a regime, diversity
  is naturally high because turnover is slow relative to innovation.

  A specific concern motivating this study was whether digital technology
  -- and in particular the visibility of popularity rankings on online
  platforms -- might accelerate the erosion of diversity through
  conformity bias. We find little evidence for this. Frequency-dependent
  selection is weak: among established tunes, popular tunes grow only
  slightly faster than their current popularity would predict under a
  neutral model ($\beta \approx 0.05$), an effect that is small relative to
  the fitness differences between individual tunes. This echoes the absence
  of progressive homogenisation in pop charts\cite{mauchEvolution2015},
  although weak conformity has been detected in other musical
  traditions\cite{youngbloodConformity2019a}, and population-level data have
  limited power to identify it\cite{acerbiConformity2016}. This does not mean that technology has no effect
  on the tradition -- our case studies demonstrate clearly that it does.
  A single YouTube video can revive a near-dormant tune. But these effects operate
  primarily through prestige bias\cite{henrichEvolution2001} rather than conformity bias: they change \textit{which}
  tunes are favoured, not whether popular tunes are favoured
  \textit{because} they are popular. 

  We should be cautious about generalising this conclusion. Our
  observation window spans roughly thirteen years of a tradition that was
  already well-established. The concern about
  technology homogenising culture\cite{andersonAlgorithmic2020,belloCultural2021} is typically about longer timescales and
  about the transition from purely oral to digitally mediated
  transmission. It is possible that such a transition occurred before our
  data begins, or that homogenising effects operate too slowly to detect
  in thirteen years. What we can say is that, within the digital
  ecosystem we observe, innovation outpaces selection. The tradition is
  not merely surviving its encounter with the internet -- it is
  diversifying through it.

\nsb{Modeling Choices.}
  Our models make simplifying choices that merit discussion. The
  neutral model assumes that tunes spread through peer exposure, with
  the probability of learning a tune proportional to its relative
  popularity. In reality, musicians encounter tunes through many
  channels -- recordings, books, online resources -- not only through
  peers at sessions. Our choice is constrained by the data: tunebook
  adds offer a proxy for the tunes that members of the community know,
  not for how they came to know them. But the choice is also
  principled. Session etiquette encourages musicians to play tunes
  that others know, so peer learning -- or at least peer-mediated
  selection of what to learn -- should dominate the dynamics of the
  active repertoire.

  A related concern is the role of recordings. We treat prestige
  recordings as a predictor of the selection coefficient, thereby
  modelling them as part of selection rather than transmission. One
  could argue that recordings affect both: a tune on a widely heard
  album receives greater exposure (a transmission effect), but a tune
  played by a master may also sound more appealing than the same tune
  played by an amateur, increasing the probability that a listener
  decides to learn it (a selection effect). This apparent ambiguity is
  not a flaw in the modelling framework -- it is a standard feature of
  evolutionary models, where fitness routinely incorporates both
  survival and reproduction without requiring that the two be
  separated\cite{orrFitness2009}. In the same way, prestige recordings can simultaneously
  increase a tune's exposure and enhance its attractiveness, and our
  models accommodate both effects through the per-tune fitness
  parameter.

\nsb{Selection for intrinsic melodic features}
  Prestige is not the whole story. ``The Bag of Potatoes''
  (tune \texttt{391}) has been recorded \num{17} times, including by
  influential artists such as The Chieftains, yet its fitness is low
  ($\beta = -0.32$) -- even lower than the XGBoost model predicts
   -- and its growth is markedly slower than that of
  comparably popular tunes. The tune is not obscure:
  at Week 0 it ranked \num{321}\textsuperscript{st} in popularity, but
  by the final week it had fallen to position \num{501}. One interpretation
  is that prestigious recordings once boosted this tune's popularity, but
  the tune itself is unable to thrive on its own merits after the
  popularity boost wears off -- prestige without sustained fitness.

  The mirror image -- fitness without obvious prestige -- is harder to
  demonstrate, as it is not clear how much data one would need to gather
  in order to rule out prestige. ``Da Lounge Bar'' (tune \texttt{8853}) is a
  young tune, composed about \num{20} years ago. It has not been widely
  recorded, and the available recordings have relatively low Spotify
  popularity, leading to a low $\beta$ prediction.
  Yet it is one of the fittest tunes in the dataset ($\beta=1.56$).
  A Youtube video from 2019 has accumulated almost 200k views, but the
  tune's growth profile appears smooth and predates the video
  , suggesting that online exposure is not the primary
  driver. In the absence of any clear prestige boost, we consider this a
  candidate for a tune whose intrinsic properties -- catchiness,
  memorability, ease of learning -- convert listening into learning at
  an unusually high rate\cite{janssenPredicting2017a}.

\nsb{Limitations.}
  Several limitations of this study should be noted. First, our data are
  drawn from a single online platform, and its users are a self-selected
  community of enthusiasts. The tunes that are popular on TheSession.org
  may not perfectly reflect the tunes that are popular in sessions, and
  the dynamics we observe are those of an online community that overlaps
  with, but is not identical to, the broader tradition. Musicians who do
  not use the website are invisible to our analysis.

  Second, the unit of observation -- a tunebook add -- is a proxy for
  learning and playing a tune, not a direct measure of it. A user saving
  a tune to their online tunebook may have just learned it, may intend to
  learn it, or may simply be bookmarking it for reference. Conversely,
  many tunes are learned and played without ever being recorded on the
  platform. The actual transmission of tunes happens at sessions,
  which we do not observe directly.

  Third, our covariates for tune fitness are coarse. Spotify popularity
  reflects listeners broadly, not the subset of musicians who drive oral
  transmission. Measures like entropy and note density capture
  aspects of melodic complexity, but the properties we would most like to
  measure -- \eg, memorability, learnability, and stylistic typicality -- are
  not reducible to summary statistics of the notated melody. The low
  variance explained by our predictive model (\SI{29}{\%}) reflects this
  gap. We note also that the ecological dynamics of tune popularity are
  inherently complex and context-dependent: even with ideal covariates,
  we would not expect to predict tune success with high accuracy.

\nsb{Irish traditional music as a model system for cultural evolution.}
  A recurring theme of this study is the gap between what we can measure
  and what we would like to measure. We can detect that tunes differ in
  fitness; we cannot yet explain why. We can show that prestige
  recordings boost specific tunes; we cannot quantify how a tune's
  intrinsic properties determine whether that boost is sustained. Closing
  this gap will require richer phenotypic descriptions, controlled
  experiments, and more detailed models of transmission -- and we argue
  that Irish traditional music is an unusually promising system in which
  to pursue all three.

  The key advantages are constraint and data richness.
  The tune's melody is the primary unit that is transmitted,
  and its phenotype is tightly bounded: most tunes are 32 bars long, conform to
  the tonality of one of four modes, and follow one of a
  small number of rhythmic types (\eg, reel, jig)\cite{dohertyMelodic2022}.
  This regularity means that meaningful comparisons
  between tunes are possible. A distance metric between two reels in the
  same key is well-defined in a way that a distance between a hip-hop
  track and a power ballad is not. This constraint 
  makes the phenotypic space navigable and makes it feasible to
  ask precise questions about which melodic features predict fitness.

  The transmission process is similarly constrained. Unlike pop music,
  where adoption can mean anything from passive streaming to karaoke to
  professional cover performance, the adoption of an Irish tune involves
  a specific sequence of steps: a musician hears the tune, decides it is
  worth learning, practises it (typically by ear), and eventually plays
  it at a session. Each step is a filter, and a tune's fitness is the
  compound probability of passing all of them. This process is concrete
  enough to be modelled and, crucially, to be tested experimentally.

  The historical depth of the tradition is a further asset. Published
  collections of Irish tunes -- O'Neill's \textit{Music of Ireland}
  (1903), Breathnach's \textit{Ceol Rince na h\'{E}ireann} (1963), and
  many others\cite{fleischmannSources1998} -- provide a fossil record extending back over two centuries.
  Tunes in our dataset can be traced through these collections, opening the
  possibility of studying fitness and selection over timescales far longer
  than the thirteen years of digital data analysed here.

  What, then, is needed to move from detecting selection to explaining
  it? We suggest three directions. First, richer phenotypic descriptions
  of tunes. Summary statistics such as entropy and note density are too
  coarse to capture what makes a tune memorable or learnable. Modern
  approaches -- including melodic embeddings derived from machine
  learning models trained on large corpora of music -- could provide
  high-dimensional phenotypic representations that better predict
  fitness. Second, controlled experiments on learning and
  recall\cite{verhoefMelodic2021a,anglada-tortLargescale2023a}.
  Presenting musicians with tunes of varying complexity or style-typicality and
  testing retention after fixed intervals would provide direct estimates
  of learnability and long-term recall -- the properties most likely to underlie intrinsic
  fitness. Memory experiments involving recognition would complement this by
  isolating memorability from motor difficulty. Third, finer-grained data on transmission.
  Tracking how tunes spread through a community of session musicians -- who
  learned what from whom, and when -- would allow the construction of
  transmission trees analogous to phylogenies in biology, enabling direct
  estimation of tune-level reproductive rates.

\small
\section*{Methods} 
\subsection*{Data}

  \textit{TheSession.org} is a popular online platform for Irish traditional music
  with \num{106000} members. Each member has a ``tunebook'', which nominally indicates
  the tunes that are known to that user. The website owner provided us with a dataset (DS1) of 
  \num{820752} individual tunebook-add events -- timestamps of when a user adds the tune to their
  tunebook -- recorded from December 2012 to March
  2026, covering \num{22216} unique tunes.

  DS1 events are aggregated into $T = 695$ fixed-width, Monday-anchored
  7-day bins covering the full DS1 date range. This results in an
  add-count matrix of shape $T \times M = 695 \times 22{,}216$, of which
  the $T - 1 = 694$ week-to-week transitions enter the model likelihood.
  We combined this with a separate dataset
  obtained from Github (DS2, Supplementary Section 1) of
  weekly total tunebook members per tune from October 2019 to March 2026. The weekly total
  accounts for both adds and removals, giving us an almost complete picture of the activity.
  We reconstruct a weekly frequency series by starting with the most recent tunebook totals,
  and back-calculating earlier values by cumulative subtraction of the weekly add counts.
  Values are clipped at zero because DS1 does not record removes.
  The main analyses use a subset (DS1b) that excludes the \num{10425}
  tunes whose reconstructed tunebook count never reaches \num{10} members,
  and any remaining week--tune cells with a count below \num{10}; this
  leaves \num{11791} tunes. Results for the full DS1 are reported in the
  Supplementary Material.

\subsection*{Data Cleaning}

  Within-tune add timestamps in DS1 were clustered using DBSCAN
  ($\varepsilon = 120$\,s, \texttt{min\_samples} $= 3$) to detect rapid-fire burst
  events that are unlikely to represent genuine independent user actions.  6 tunes
  have a burst fraction $\geq 0.5$, meaning at least \SI{50}{\%} of their lifetime add
  events fall within burst clusters. Burst events for these tunes are removed
  from DS1 before interval aggregation (63 events); all other adds are retained.
  The number of flagged tunes is insensitive to $\varepsilon$ over 30--300\,s
  once \texttt{min\_samples} $\geq 3$.

\subsection*{Models}
\label{sec:models}

\subsubsection*{Overview}

  Tune popularity is modelled as a discrete-time birth process operating
  on a pool of tunes.  At each timestep $t$, tune $i$ has an absolute tunebook
  count $\Fti$ (the number of registered members who have added it) and a
  normalised relative popularity
  \begin{equation}
      \fti = \frac{\Fti}{\displaystyle\sum_{j:\,F_{t,j}>0} F_{t,j}},
      \label{eq:fti}
  \end{equation}
  where the sum runs over all tunes active at time $t$ (\ie, with $F_{t,j}>0$).
  Cells with $\Fti = 0$ are treated as inactive and are skipped during
  likelihood evaluation. We primarily report results of a modelling framework
  where births are modelled using a multinomial distribution.
  We implemented additional modelling frameworks using DS2, including a Skellam
  process\cite{skellamFrequency1946} that models both births and deaths (Supplementary Section).
  These analyses confirm that our methodology produces robust results, and that
  the effects of tune removals -- present in DS2 but not DS1 -- are negligible.

  All models assign a weight $\wti$ to each active tune and derive
  selection probabilities by normalisation:
  \begin{equation}
      \pti = \frac{\wti}{\displaystyle\sum_{j:\,F_{t,j}>0} w_{t,j}}.
      \label{eq:pti}
  \end{equation}

  Add events in DS1 are modelled as independent draws from a
  categorical distribution over active tunes:
  \begin{equation}
      \ell^{\mathrm{mult}} = \sum_{t}\ \sum_{i:\,F_{t,i}>0} \atti \log \pti,
      \label{eq:multinomial}
  \end{equation}
  where $\atti$ is the number of adds of tune $i$ at time $t+1$, and the
  outer sum runs over all adjacent pairs of timesteps.
  One neutral and two selection models are considered,
  differing in how $\wti$ depends on $\fti$.

  \paragraph{Neutral Model.}
  Since we are using a population ecology framework in a cultural setting, it is important
  to explain the logic behind the model -- we are not naively transposing one modelling 
  framework across domains without considering the meaning behind the mathematics.
  The baseline model of neutral evolution used here is similar to a Wright-Fisher model:
  new tunebook adds at time $t + 1$ are assumed to be sampled with replacement from
  the pool of tunebook entries at time $t$.
  \begin{equation}
      \wti = \fti.
  \end{equation}
  This is consistent with a model where musicians learn tunes with a probability proportional
  to the rate at which they encounter the tune from their peers; this is equivalent to
  a neutral birth process where all parents have equal chance of reproductive success.
  In reality populations are structured -- peer groups have a network structure -- but we do
  not have information on this so equate the probability of encountering a tune to its
  relative popularity -- this is equivalent to the Wright Fisher model assumption of random mating.
  The DS1 dataset contains no removals -- hence no death or extinction -- but there
  are events for the inception of a new tune -- analogous to immigration or speciation.
  We include these new tunes once they are present, but we do not explicitly model the inception event.

  \paragraph{Frequency-dependent selection.}
  A single global parameter $\sbeta$ modulates whether popular tunes are
  disproportionately likely to be added ($\sbeta > 0$, conformity bias)
  or suppressed ($\sbeta < 0$, novelty bias):
  \begin{equation}
      \wti = \fti^{1+\sbeta}.
      \label{eq:freqdep}
  \end{equation}

  \paragraph{Per-tune selection.}
  Each tune $i$ carries its own fixed fitness offset $\sbeta_i$:
  \begin{equation}
      \wti = \fti \cdot \exp(\sbeta_i).
      \label{eq:pertune}
  \end{equation}

  \paragraph{Per-tune selection with saturation.}
  The unsaturated per-tune model allows the expected adds for tune $i$ to grow
  without bound, but in practice every user can add a given tune at most once,
  so the effective pool of users who could still add tune $i$ shrinks as
  $\Fti$ grows.  We capture this by multiplying the per-tune weight by a
  susceptible-fraction factor analogous to the $(1 - I/N)$ term in
  compartmental epidemic models:
  \begin{equation}
      \wti = \fti \cdot \exp(\sbeta_i) \cdot (1 - \sval\,\Fti),
      \qquad \sval = \smax\,\sigma(\sgamma),
      \label{eq:pertune_sat}
  \end{equation}
  where $\sigma$ is the logistic function, $\smax = 1/\max_{t,i} \Fti$ ensures
  the saturation factor stays in $(0,1)$, and $\sgamma$ is a single
  global free parameter fitted jointly with the $\sbeta_i$.  At the fitted
  $\sgamma$, the inverse $1/\sval$ has the interpretation of an effective
  user pool: the number of users available to add any given tune.  For DS1
  the fitted value implies an effective pool of approximately \num{22500}
  users.  This is biologically plausible: although TheSession.org has
  \num{106000} registered members, only about \num{54000} have ever added any
  tune to their tunebook (personal communication from the site owner), and
  only a fraction of those are actively adding tunes during any given period.
  Equation~\eqref{eq:pertune_sat} is the default model used to obtain the
  per-tune fitness estimates reported throughout the paper; results for the
  unsaturated form in \eqref{eq:pertune} are virtually identical for the vast
  majority of tunes and are reported in the Supplementary Material.

  \subsubsection*{Ridge Regularisation}
  \label{sec:ridge}

  All models with free parameters are fitted under a Gaussian ridge prior placed
  independently on each parameter $\beta_j$:
  \begin{equation}
      \log \pi(\boldsymbol{\beta}) = -\frac{\|\boldsymbol{\beta}\|^2}{2\sridge^2} + \mathrm{const},
      \label{eq:ridge}
  \end{equation}
  with default prior standard deviation $\sridge = 0.2$ (for the saturation
  models the prior is placed on $\sval$ rather than on $\sgamma$).  The objective
  minimised during optimisation is the negative log-posterior,
  \begin{equation}
      \mathcal{L}(\boldsymbol{\beta}) = -\ell(\boldsymbol{\beta}) + \frac{\|\boldsymbol{\beta}\|^2}{2\sridge^2},
      \label{eq:objective}
  \end{equation}
  where $\ell$ is the log-likelihood \eqref{eq:multinomial}.
  The ridge prior prevents overfitting to sparse data in cases where there are few add events.
  A sensitivity analysis of the inferred $\sbeta$ is reported in Supplementary Fig.~S5.

  \subsubsection*{Optimisation}
  \label{sec:optimisation}

  The objective \eqref{eq:objective} is minimised using the limited-memory BFGS
  (L-BFGS) algorithm\cite{liuLimited1989} with a strong Wolfe line-search condition.
  Analytic gradients $\nabla_{\boldsymbol{\beta}}
  \mathcal{L}$ are computed for all selection models, enabling efficient
  high-dimensional optimisation.  
  The optimiser is run for at most 500 L-BFGS iterations; convergence is
  declared when the gradient norm falls below $10^{-2}$.

  \subsubsection*{Uncertainty Quantification}
  \label{sec:uq}

  Standard errors for the MAP estimates $\boldsymbol{\beta}$ are obtained
  under the Laplace approximation, which approximates the posterior as a
  Gaussian with covariance equal to the inverse Hessian of $\mathcal{L}$ at
  $\boldsymbol{\beta}$.  Because the off-diagonal Hessian elements are not
  required for per-parameter inference, only the diagonal is computed, using
  central finite differences:
  \begin{equation}
      H_{jj} \approx \frac{\mathcal{L}(\boldsymbol{\beta} + \epsilon\,\mathbf{e}_j)
                         - 2\mathcal{L}(\boldsymbol{\beta})
                         + \mathcal{L}(\boldsymbol{\beta} - \epsilon\,\mathbf{e}_j)}{\epsilon^2},
      \label{eq:hessian}
  \end{equation}
  where $\mathbf{e}_j$ is the $j$-th standard basis vector and $\epsilon$ is a
  small step size (default $10^{-4}$). The marginal
  posterior standard error is then $\mathrm{SE}_j = 1/\sqrt{H_{jj}}$.  If a
  diagonal element is non-positive (indicating a numerically flat or
  non-convex curvature), the SE falls back to the prior standard deviation
  $\sridge$ as a conservative upper bound.

  For each parameter, a Wald $z$-statistic and two-tailed $p$-value are computed
  as $z_j = \beta_j / \mathrm{SE}_j$ and $p_j = 2\Phi(-|z_j|)$, where
  $\Phi$ is the standard normal CDF.  For the per-tune selection model,
  the $p$-values are additionally corrected for multiple comparisons using the
  Benjamini-Hochberg false-discovery-rate procedure\cite{benjaminiControlling1995} as implemented in
  \texttt{scipy}. 

  \subsubsection*{Model Comparison}
  \label{sec:aic}

  Models are compared using the Akaike Information Criterion (AIC)\cite{akaikeNew1974}:
  \begin{equation}
      \mathrm{AIC} = 2k - 2\ell,
      \label{eq:aic}
  \end{equation}
  where $k$ is the number of free parameters and $\ell$ is the ridge-free
  log-likelihood at the MAP estimate.  The neutral model has no free
  parameters; the frequency-dependent model has one ($\sbeta$); the
  per-tune model has one per tune ($k = 11{,}791$ for DS1b); the saturated variants
  each carry one additional global parameter ($\sgamma$).

  \subsubsection*{Sliding-Window Fitness Trajectories}
  \label{sec:sliding_window}

  To examine temporal variation in per-tune fitness, the saturated per-tune model
  is applied repeatedly to overlapping windows of the DS1 time series.
  Each window spans 2 years (104 weekly timesteps) and windows are advanced in
  steps of 0.5 years (26 timesteps), yielding a sequence of 23
  overlapping-window $\beta_i$ estimates.  Within each window the same ridge prior
  ($\sigma = 0.2$) and L-BFGS optimiser are used as for the full-data fit;
  all $\beta_i$ are free parameters, and tunes with no add events within the
  window are held at $\beta_i = 0$ by the prior. Standard errors are not
  computed for windowed fits, and estimates are plotted at the end of
  their window.

  \subsection*{Popularity Covariates}

  \subsubsection*{Recording Prestige Proxies}
  The Github repository for TheSession.org contains detailed user-contributed
  data mapping tunes to individual tracks on commercial recordings. These
  were used to map to information on Discogs and Spotify using their respective APIs.

  \paragraph{TheSession recordings.}
  We count the number of recorded tracks on which a tune appears. Since the count distribution
  is long-tailed, and has many zero values, we use the $\log_{1\mathrm{p}}$-transformed
  counts as a feature in the predictive model.

  \paragraph{Discogs.}
  We searched on Discogs for recordings identified by users on TheSession.org,
  and counted the number of users on Discogs that indicated that they \textit{have}
  that recording. We assigned a count of zero to recordings that could not be
  matched on Discogs. For each tune we sum have-counts over all recordings that
  a tune is on. Since the count distribution is long-tailed, and has many zero
  values, we use the $\log_{1\mathrm{p}}$-transformed sum-have-counts as a feature
  in the predictive model. This feature is less precise than the Spotify measure,
  since it is measured at the level of a recording, whereas the Spotify measure
  is at the track level.

  \paragraph{Spotify.}
  TheSession recordings are matched to Spotify album JSON data via fuzzy
  album-name matching (\texttt{SequenceMatcher} ratio $\geq 0.85$, track count
  agreement within $\pm 3$ tracks).  For each matched track the Spotify platform
  popularity score (an integer on \SIrange{0}{100} reflecting recent streaming activity) is
  retrieved. For each tune we note the maximum popularity; unmatched tracks are
  assigned a popularity of zero. Since the max-popularity distribution is long-tailed, and has many zero
  values, we use the $\log_{1\mathrm{p}}$-transformed values as features
  in the predictive model.

  \paragraph{Additional proxies computed but excluded.}
  Discogs wish-list count and log mean / median Spotify popularity
  were also computed but excluded from the headline analysis
  after preliminary screening: Discogs wish-list counts is
  highly collinear with Discogs have-counts, as are the mean/median Spotify
  scores with the max Spotify score.

  \subsubsection*{Intrinsic Melodic Features}

  Five measures of melodic structure are computed from the ABC notation for each
  tune setting and then aggregated to the tune level by taking the \emph{median} across
  all settings for that tune.  The measures are:

  \begin{itemize}
      \item \textbf{Melodic Pitch-class Entropy} --  $H = -\sum_k p_k \log_2 p_k$ of
            the pitch-class sequence.

      \item \textbf{Note density} -- total note onsets divided by the tune length
            in units of quarter notes. Onsets for unison (non-changing) intervals
            are not considered. In the Irish tradition, long notes and unison
            pairs or triplets are exchangeable, and essentially uninformative.

      \item \textbf{LZ76} -- Lempel-Ziv 1976 complexity\cite{lempelComplexity1976} of the
            pitch class sequence, normalised as $c(n)\log_2 n\,/\,(n\log_2 k)$ with
            fixed alphabet $k = 12$, where $c(n)$ is the raw phrase count produced
            by the Kaspar--Schuster greedy exhaustive parsing algorithm and $n$ is
            the sequence length.  This normalisation converges to 1 for i.i.d.\
            uniform sequences and yields values $< 1$ for repetitive sequences.

      \item \textbf{Mean $|$interval$|$} -- mean absolute size (in semitones) of
            all non-zero melodic intervals.

      \item \textbf{Contour complexity} -- proportion of direction changes among
            consecutive non-zero intervals. A value of 0 indicates
            a monotonic melody; 1 indicates alternating up-down motion.
  \end{itemize}

  A wider set of melodic descriptors (the absolute deviation from corpus-mean
  entropy, MIDI-pitch entropy, number of bars, raw note density including
  unisons, total number of notes, pitch range, number of distinct pitches, and
  LZ76 complexity of the interval sequence) was also computed but excluded from
  the headline analysis after preliminary screening, on the basis of high
  collinearity with the retained measures and negligible marginal contribution
  to predictive accuracy.

  \subsection*{Modelling Tune Fitness}
  \label{sec:covariate_analysis}

  \subsubsection*{XGBoost predictor of $\beta_i$}

  We model each tune's fitness $\beta_i$ as a non-linear function of nine
  covariates -- three prestige proxies, five melodic
  features, and a single popularity baseline
  $\log \bar F_i$, where $\bar F_i = \frac{1}{T}\sum_t F_{t,i}$ is the
  time-averaged tunebook count of tune $i$.  The prestige and melodic covariates
  are $z$-scored before fitting.  Analyses are restricted to tunes with
  $\bar F_i \geq 10$, for which $\beta_i$ is reliably estimated (Supplementary
  Fig.~S2); tunes with any missing covariate are dropped, yielding
  $n = 8051$ matched tunes.

  The regressor is a gradient-boosted decision-tree ensemble\cite{friedmanGreedy2001,chenXgboost2016}
  fitted to minimise weighted squared error,
  with per-tune weights $w_i = 1/\mathrm{SE}_i^2$ taken from the Laplace
  approximation.  Hyperparameters are deliberately
  conservative to keep the in-sample / cross-validation gap small:
  \texttt{max\_depth} $= 3$, \texttt{min\_child\_weight} $= 200$,
  \texttt{learning\_rate} $= 0.03$, \texttt{n\_estimators} $= 300$,
  \texttt{subsample} $= 0.8$, \texttt{colsample\_bytree} $= 0.8$, with default
  $\ell_2$ regularisation ($\lambda = 1$).  Out-of-fold predictions $\hat\beta_i$
  are obtained from 5-fold cross-validation with a fixed fold assignment (seed
  42); the in-sample $R^2$ (0.42) exceeds the CV $R^2$ (0.29) by 0.13.  This same fit
  is used for the predicted values shown in the Supplementary Figure.

  \subsubsection*{Variance Partitioning}

  To attribute predictive accuracy to covariate groups, we re-run the same
  XGBoost configuration on (i) each group in isolation and (ii) the full
  model with that group removed.  The same fold assignment is used across all
  fits within one call so the marginal and leave-one-group-out $R^2$ values
  are directly comparable.  For each group $g$ we report
  \begin{align}
      R^2_\mathrm{marginal}(g) &= R^2_\mathrm{CV}\!\left(\text{model on group }g\text{ only}\right),\\
      R^2_\mathrm{unique}(g)   &= R^2_\mathrm{full} - R^2_\mathrm{CV}\!\left(\text{full model with group }g\text{ removed}\right),
  \end{align}
  with $R^2_\mathrm{full}$ the 5-fold CV $R^2$ of the model using all groups.
  The marginal value reflects total signal in the group; the unique value
  isolates the portion of that signal not redundant with the other groups.
  All $R^2$ values are sample-weighted by $w_i$.
  
  \subsubsection*{GAM Partial Effects}
  \label{sec:gam}

  To visualise the direction and shape of each covariate's association with
  $\beta_i$, we fit a generalised additive model (GAM)\cite{woodGeneralized2017} with one penalised
  spline smooth per covariate (\texttt{pygam.LinearGAM}; 10 splines per
  term, smoothing penalty $\lambda = 10$), using the same nine $z$-scored
  covariates, $1/\mathrm{SE}_i^2$ sample weights, and $n = 8051$ tunes as
  the XGBoost model.  \fref{fig:fig2}b shows the fitted smooths for four
  predictors as partial effects on $\beta_i$ with \SI{95}{\%} confidence
  intervals, drawn over the 1st--99th percentiles of each covariate.
  Unlike the tree ensemble, the GAM is additive by construction, so each
  smooth is directly interpretable as a marginal effect; the corresponding
  XGBoost partial-dependence curves (Supplementary Material) show the same
  qualitative pattern.

  \subsection*{Set Co-occurrence Analysis}
  \label{sec:cooccurrence}

  Irish traditional music is typically performed in \textit{sets} — sequences of
  two or more tunes played consecutively.  TheSession records user-submitted
  sets in the Github data repository, dating from 2016.

  \subsubsection*{Frequency of Recorded Sets}

  We restrict attention to the \num{41743} recorded sets of exactly two
  tunes, treating each as an unordered combination of tune identifiers.
  For each distinct combination we count the number of times $c$ it is
  recorded, and report the number of distinct combinations recorded
  exactly $c$ times (\fref{fig:fig2}d).

  \subsubsection*{Null Model}

  Under a popularity-only null, the same number of two-tune sets is
  assembled by drawing tunes sequentially without replacement, with
  probability proportional to each tune's time-averaged share of
  \texttt{tunebook} entries in DS1. The frequency-of-frequencies
  distribution is averaged over 20 simulated corpora. Under this null,
  \num{40788} of the \num{41743} sets are unique on average and no
  combination is recorded more than 8 times, compared with \num{22982}
  distinct combinations and a maximum of 134 recordings in the observed
  data.

\bibliography{TunePopularity}

\begin{thebibliography}{43}
\providecommand{\natexlab}[1]{#1}
\providecommand{\url}[1]{\texttt{#1}}
\expandafter\ifx\csname urlstyle\endcsname\relax
  \providecommand{\doi}[1]{doi: #1}\else
  \providecommand{\doi}{doi: \begingroup \urlstyle{rm}\Url}\fi

\bibitem[Cavalli-Sforza and Feldman(1981)]{cavallisforzaCultural1981}
Luigi~Luca Cavalli-Sforza and Marcus~W. Feldman.
\newblock \emph{Cultural Transmission and Evolution: A Quantitative Approach}.
\newblock Princeton University Press, Princeton, NJ, 1981.

\bibitem[Boyd and Richerson(1988)]{boydCulture1988}
Robert Boyd and Peter~J. Richerson.
\newblock \emph{Culture and the Evolutionary Process}.
\newblock University of Chicago Press, Chicago, paperback ed edition, 1988.

\bibitem[Henrich and Boyd(1998)]{henrichEvolution1998}
Joe Henrich and Robert Boyd.
\newblock The evolution of conformist transmission and the emergence of
  between-group differences.
\newblock \emph{Evolution and Human Behavior}, 19\penalty0 (4):\penalty0
  215--241, 1998.
\newblock \doi{10.1016/s1090-5138(98)00018-x}.

\bibitem[Hahn and Bentley(2003)]{hahnDrift2003}
Matthew~W. Hahn and R.~Alexander Bentley.
\newblock Drift as a mechanism for cultural change: an example from baby names.
\newblock \emph{Proceedings of the Royal Society of London. Series B:
  Biological Sciences}, 270\penalty0 (Suppl. 1), 2003.
\newblock \doi{10.1098/rsbl.2003.0045}.

\bibitem[Bentley et~al.(2004)Bentley, Hahn, and Shennan]{bentleyRandom2004}
R.~Alexander Bentley, Matthew~W. Hahn, and Stephen~J. Shennan.
\newblock Random drift and culture change.
\newblock \emph{Proc. R. Soc. Lond. B Biol. Sci.}, 271\penalty0
  (1547):\penalty0 1443--1450, 2004.
\newblock \doi{10.1098/rspb.2004.2746}.

\bibitem[Bentley et~al.(2007)Bentley, Lipo, Herzog, and
  Hahn]{bentleyRegular2007}
R.~Alexander Bentley, Carl~P. Lipo, Harold~A. Herzog, and Matthew~W. Hahn.
\newblock Regular rates of popular culture change reflect random copying.
\newblock \emph{Evol. Hum. Behav.}, 2007.
\newblock \doi{10.1016/j.evolhumbehav.2006.10.002}.

\bibitem[Shennan and Wilkinson(2001)]{shennanCeramic2001}
S.~J. Shennan and J.~R. Wilkinson.
\newblock Ceramic style change and neutral evolution: A case study from
  neolithic europe.
\newblock \emph{American Antiquity}, 66\penalty0 (4):\penalty0 577--593, 2001.
\newblock \doi{10.2307/2694174}.

\bibitem[Herzog et~al.(2004)Herzog, Bentley, and Hahn]{herzogRandom2004}
H.~A. Herzog, R.~A. Bentley, and M.~W. Hahn.
\newblock Random drift and large shifts in popularity of dog breeds.
\newblock \emph{Proceedings of the Royal Society of London. Series B:
  Biological Sciences}, 271\penalty0 (Suppl. 5), 2004.
\newblock \doi{10.1098/rsbl.2004.0185}.

\bibitem[Newberry et~al.(2017)Newberry, Ahern, Clark, and
  Plotkin]{newberryDetecting2017}
Mitchell~G. Newberry, Christopher~A. Ahern, Robin Clark, and Joshua~B. Plotkin.
\newblock Detecting evolutionary forces in language change.
\newblock \emph{Nature}, 551\penalty0 (7679):\penalty0 223--226, 2017.
\newblock \doi{10.1038/nature24455}.

\bibitem[Youngblood(2019)]{youngbloodConformity2019a}
Mason Youngblood.
\newblock Conformity bias in the cultural transmission of music sampling
  traditions.
\newblock \emph{R. Soc. Open Sci.}, 6\penalty0 (9):\penalty0 191149, 2019.
\newblock \doi{10.1098/rsos.191149}.

\bibitem[Kandler and Crema(2019)]{kandlerAnalysing2019}
Anne Kandler and Enrico~R. Crema.
\newblock Analysing cultural frequency data: Neutral theory and beyond.
\newblock In \emph{Handbook of Evolutionary Research in Archaeology}, pages
  83--108. Springer International Publishing, 2019.
\newblock \doi{10.1007/978-3-030-11117-5_5}.

\bibitem[Leroi et~al.(2020)Leroi, Lambert, Rosindell, Zhang, and
  Kokkoris]{leroiNeutral2020}
Armand~M. Leroi, Ben Lambert, James Rosindell, Xiangyu Zhang, and Giorgos~D.
  Kokkoris.
\newblock Neutral syndrome.
\newblock \emph{Nat. Hum. Behav.}, 2020.
\newblock \doi{10.1038/s41562-020-0844-7}.

\bibitem[Breathnach(1986)]{breathnachUse1986}
Breand{\'a}n Breathnach.
\newblock \emph{The Use of Notation in the Transmission of {{Irish}} Folk
  Music}.
\newblock \'O {{Riada}} Memorial Lecture 1. Irish Traditional Music Society,
  University College Cork, Cork, 1986.

\bibitem[Cawley(2020)]{cawleyBecoming2020}
Jessica Cawley.
\newblock \emph{Becoming an Irish Traditional Musician}.
\newblock Routledge, 2020.
\newblock \doi{10.4324/9781003083344}.

\bibitem[O’Shea(2007)]{osheaGetting2007}
Helen O’Shea.
\newblock Getting to the heart of the music: Idealizing musical community and
  irish traditional music sessions.
\newblock \emph{Journal of the Society for Musicology in Ireland}, pages 1--18,
  2007.
\newblock \doi{10.35561/jsmi02061}.

\bibitem[Ward(2019)]{wardTechnology2019}
Francis Ward.
\newblock Technology and the transmission of tradition: {{An}} exploration of
  the virtual pedagogies in the {{Online Academy}} of {{Irish Music}}.
\newblock \emph{J. Music Technol. Amp Educ.}, 2019.
\newblock \doi{10.1386/jmte.12.1.5_1}.

\bibitem[Walshaw()]{walshawAbcnotationcom}
Chris Walshaw.
\newblock {{ABCNotation}}.
\newblock URL \url{abcnotation.com}.

\bibitem[Keith()]{keithThesessionorg}
Jeremy Keith.
\newblock The {{Session}}.
\newblock URL \url{thesession.org}.

\bibitem[Street et~al.(2022)Street, Eerola, and Kendal]{streetRole2022}
Sally~E. Street, Tuomas Eerola, and Jeremy~R. Kendal.
\newblock The role of population size in folk tune complexity.
\newblock \emph{Humanit. Soc. Sci. Commun.}, 9\penalty0 (1):\penalty0 152,
  2022.
\newblock \doi{10.1057/s41599-022-01139-y}.

\bibitem[Condit et~al.(2012)Condit, Chisholm, and Hubbell]{conditThirty2012}
Richard Condit, Ryan~A. Chisholm, and Stephen~P. Hubbell.
\newblock Thirty years of forest census at barro colorado and the importance of
  immigration in maintaining diversity.
\newblock \emph{PLoS ONE}, 7\penalty0 (11):\penalty0 e49826, 2012.
\newblock \doi{10.1371/journal.pone.0049826}.

\bibitem[Barton(2000)]{bartonGenetic2000}
N.~H. Barton.
\newblock Genetic hitchhiking.
\newblock \emph{Philosophical Transactions of the Royal Society of London.
  Series B: Biological Sciences}, 355\penalty0 (1403):\penalty0 1553--1562,
  2000.
\newblock \doi{10.1098/rstb.2000.0716}.

\bibitem[Chesson(2000)]{chessonMechanisms2000}
Peter Chesson.
\newblock Mechanisms of maintenance of species diversity.
\newblock \emph{Annual Review of Ecology and Systematics}, 31\penalty0
  (1):\penalty0 343--366, 2000.
\newblock \doi{10.1146/annurev.ecolsys.31.1.343}.

\bibitem[Whitlock(2003)]{whitlockFixation2003}
Michael~C Whitlock.
\newblock Fixation probability and time in subdivided populations.
\newblock \emph{Genetics}, 164\penalty0 (2):\penalty0 767--779, 2003.
\newblock \doi{10.1093/genetics/164.2.767}.

\bibitem[Ohta(1992)]{ohtaNearly1992}
Tomoko Ohta.
\newblock The nearly neutral theory of molecular evolution.
\newblock \emph{Annual Review of Ecology and Systematics}, 23\penalty0
  (1):\penalty0 263--286, 1992.
\newblock \doi{10.1146/annurev.es.23.110192.001403}.

\bibitem[Mauch et~al.(2015)Mauch, MacCallum, Levy, and
  Leroi]{mauchEvolution2015}
Matthias Mauch, Robert~M. MacCallum, Mark Levy, and Armand~M. Leroi.
\newblock The evolution of popular music: {{USA}} 1960--2010.
\newblock \emph{R. Soc. Open Sci.}, 2\penalty0 (5):\penalty0 150081, 2015.
\newblock \doi{10.1098/rsos.150081}.

\bibitem[Acerbi et~al.(2016)Acerbi, van Leeuwen, Haun, and
  Tennie]{acerbiConformity2016}
Alberto Acerbi, Edwin J.~C. van Leeuwen, Daniel B.~M. Haun, and Claudio Tennie.
\newblock Conformity cannot be identified based on population-level signatures.
\newblock \emph{Scientific Reports}, 6\penalty0 (1):\penalty0 36068, 2016.
\newblock \doi{10.1038/srep36068}.

\bibitem[Henrich and Gil-White(2001)]{henrichEvolution2001}
Joseph Henrich and Francisco~J Gil-White.
\newblock The evolution of prestige: freely conferred deference as a mechanism
  for enhancing the benefits of cultural transmission.
\newblock \emph{Evolution and Human Behavior}, 22\penalty0 (3):\penalty0
  165--196, 2001.
\newblock \doi{10.1016/s1090-5138(00)00071-4}.

\bibitem[Anderson et~al.(2020)Anderson, Maystre, Anderson, Mehrotra, and
  Lalmas]{andersonAlgorithmic2020}
Ashton Anderson, Lucas Maystre, Ian Anderson, Rishabh Mehrotra, and Mounia
  Lalmas.
\newblock Algorithmic effects on the diversity of consumption on spotify.
\newblock In \emph{Proceedings of The Web Conference 2020}, pages 2155--2165.
  ACM, 2020.
\newblock \doi{10.1145/3366423.3380281}.

\bibitem[Bello and Garcia(2021)]{belloCultural2021}
Pablo Bello and David Garcia.
\newblock Cultural divergence in popular music: the increasing diversity of
  music consumption on spotify across countries.
\newblock \emph{Humanities and Social Sciences Communications}, 8\penalty0
  (1):\penalty0 182, 2021.
\newblock \doi{10.1057/s41599-021-00855-1}.

\bibitem[Orr(2009)]{orrFitness2009}
H.~Allen Orr.
\newblock Fitness and its role in evolutionary genetics.
\newblock \emph{Nature Reviews Genetics}, 10\penalty0 (8):\penalty0 531--539,
  2009.
\newblock \doi{10.1038/nrg2603}.

\bibitem[Janssen et~al.(2017)Janssen, Burgoyne, and
  Honing]{janssenPredicting2017a}
Berit Janssen, John~A. Burgoyne, and Henkjan Honing.
\newblock Predicting {{Variation}} of {{Folk Songs}}: {{A Corpus Analysis
  Study}} on the {{Memorability}} of {{Melodies}}.
\newblock \emph{Front. Psychol.}, 8:\penalty0 621, 2017.
\newblock \doi{10.3389/fpsyg.2017.00621}.

\bibitem[Doherty(2022)]{dohertyMelodic2022}
Se{\'a}n Doherty.
\newblock Melodic {{Structures}} in the {{Double Jigs}} of {{O}}'{{Neill}}'s
  {{{\emph{The Dance Music}}}}{\emph{ of }}{{{\emph{Ireland}}}}{\emph{: 1001
  }}{{{\emph{Gems}}}} (1907).
\newblock \emph{J. Soc. Musicol. Irel.}, pages 19--45, 2022.
\newblock \doi{10.35561/JSMI17222}.

\bibitem[Fleischmann(1998)]{fleischmannSources1998}
Aloys Fleischmann.
\newblock \emph{Sources of {{Irish}} Traditional Music, c. 1600-1855}.
\newblock Number vol. 1296 in Garland Reference Library of the Humanities.
  Garland, New York, 1998.

\bibitem[Verhoef and Ravignani(2021)]{verhoefMelodic2021a}
Tessa Verhoef and Andrea Ravignani.
\newblock Melodic {{Universals Emerge}} or {{Are Sustained Through Cultural
  Evolution}}.
\newblock \emph{Front. Psychol.}, 12:\penalty0 668300, 2021.
\newblock \doi{10.3389/fpsyg.2021.668300}.

\bibitem[{Anglada-Tort} et~al.(2023){Anglada-Tort}, Harrison, Lee, and
  Jacoby]{anglada-tortLargescale2023a}
Manuel {Anglada-Tort}, Peter~M.C. Harrison, Harin Lee, and Nori Jacoby.
\newblock Large-scale iterated singing experiments reveal oral transmission
  mechanisms underlying music evolution.
\newblock \emph{Curr. Biol.}, 33\penalty0 (8):\penalty0 1472--1486.e12, 2023.
\newblock \doi{10.1016/j.cub.2023.02.070}.

\bibitem[Skellam(1946)]{skellamFrequency1946}
J.~G. Skellam.
\newblock The frequency distribution of the difference between two poisson
  variates belonging to different populations.
\newblock \emph{Journal of the Royal Statistical Society}, 109\penalty0
  (3):\penalty0 296, 1946.
\newblock \doi{10.2307/2981372}.

\bibitem[Liu and Nocedal(1989)]{liuLimited1989}
Dong~C. Liu and Jorge Nocedal.
\newblock On the limited memory bfgs method for large scale optimization.
\newblock \emph{Mathematical Programming}, 45\penalty0 (1-3):\penalty0
  503--528, 1989.
\newblock \doi{10.1007/bf01589116}.

\bibitem[Benjamini and Hochberg(1995)]{benjaminiControlling1995}
Yoav Benjamini and Yosef Hochberg.
\newblock Controlling the {{False Discovery Rate}}: {{A Practical}} and
  {{Powerful Approach}} to {{Multiple Testing}}.
\newblock \emph{J. R. Stat. Soc. Ser. B Stat. Methodol.}, 57\penalty0
  (1):\penalty0 289--300, 1995.
\newblock \doi{10.1111/j.2517-6161.1995.tb02031.x}.

\bibitem[Akaike(1974)]{akaikeNew1974}
H.~Akaike.
\newblock A new look at the statistical model identification.
\newblock \emph{IEEE Transactions on Automatic Control}, 19\penalty0
  (6):\penalty0 716--723, 1974.
\newblock \doi{10.1109/tac.1974.1100705}.

\bibitem[Lempel and Ziv(1976)]{lempelComplexity1976}
A.~Lempel and J.~Ziv.
\newblock On the complexity of finite sequences.
\newblock \emph{IEEE Transactions on Information Theory}, 22\penalty0
  (1):\penalty0 75--81, 1976.
\newblock \doi{10.1109/tit.1976.1055501}.

\bibitem[Friedman(2001)]{friedmanGreedy2001}
Jerome~H. Friedman.
\newblock Greedy function approximation: A gradient boosting machine.
\newblock \emph{The Annals of Statistics}, 29\penalty0 (5), 2001.
\newblock \doi{10.1214/aos/1013203451}.

\bibitem[Chen and Guestrin(2016)]{chenXgboost2016}
Tianqi Chen and Carlos Guestrin.
\newblock Xgboost.
\newblock In \emph{Proceedings of the 22nd ACM SIGKDD International Conference
  on Knowledge Discovery and Data Mining}, pages 785--794. ACM, 2016.
\newblock \doi{10.1145/2939672.2939785}.

\bibitem[Wood(2017)]{woodGeneralized2017}
Simon~N. Wood.
\newblock \emph{Generalized Additive Models}.
\newblock Chapman and Hall/CRC, 2017.
\newblock \doi{10.1201/9781315370279}.

\end{thebibliography}
\bibliographystyle{unsrtnat}

\end{document}